\documentclass[a4paper,scriptaddress,twocolumn,prl,showkeys]{revtex4}
\usepackage{amsmath}
\usepackage{makeidx}
\usepackage{amsfonts}
\usepackage{graphicx,epstopdf}
\usepackage[ansinew]{inputenc}
\usepackage[usenames,dvipsnames]{pstricks}
\usepackage{subfigure}
\usepackage{pdfpages}
\usepackage{epsfig}
\usepackage{pst-grad}
\usepackage{pst-plot}
\usepackage[colorlinks,hyperindex]{hyperref}
\usepackage[dvips]{geometry}

\hypersetup{
colorlinks,
citecolor=blue,
linkcolor=black,
urlcolor=black,
}

\DeclareMathOperator{\sech}{sech}
\makeindex

\begin{document}

\title{Coupled scalar fields Oscillons and Breathers in some Lorentz
Violating Scenarios}
\author{R. A. C. Correa$^{1,}$\footnote{%
rafael.couceiro@ufabc.edu.br} and A. de Souza Dutra$^{2,}$\footnote{%
dutra@feg.unesp.br}\medskip }
\affiliation{$^{1}$ CCNH, Universidade Federal do ABC, 09210-580, Santo André, SP,
Brazil\medskip \\
$^{2}$ UNESP-Campus de Guaratinguetá, 12516-410, Guaratinguetá, SP,
Brazil\medskip }
\date{February 23, 2015}

\begin{abstract}
In this work we discuss the impact of the breaking of the Lorentz symmetry
on the usual oscillons, the so-called flat-top oscillons, and the breathers.
Our analysis is performed by using a Lorentz violation scenario rigorously
derived in the literature. We show that the Lorentz violation is responsible
for the origin of a kind of deformation of the configuration, where the
field configuration becomes oscillatory in a localized region near its
maximum value. Furthermore, we show that the Lorentz breaking symmetry
produces a displacement of the oscillon along the spatial direction, the
same feature is present in the case of breathers. We also show that the
effect of a Lorentz violation in the flat-top oscillon solution is
responsible by the shrinking of the flat-top. Furthermore, we find
analytically the outgoing radiation, this result indicates that the
amplitude of the outgoing radiation is controlled by the Lorentz breaking
parameter, in such away that this oscillon becomes more unstable than its
symmetric counterpart, however, it still has a long living nature.
\end{abstract}

\maketitle

\section{1. Introduction}

The study of nonlinear systems is becoming an area of increasing interest
along the last few decades \cite{G. B. Whitham, A. C. Scott}. In fact, such
nonlinear behavior of physical systems is found in a broad part of physical
systems nowadays. This includes condensed matter systems, field theoretical
models, modern cosmology and a large number of other domains of the physical
science \cite{R. Rajaraman}-\cite{salamanca}. One of the reasons of this
increasing interest is the fact that many of those systems present a
countable number of distinct degenerate minimal energy configurations. In
many cases that degenerate structure can be studied through simple models of
scalar fields possessing a potential with two or more degenerate minima. For
instance, in two or more spatial dimensions, one can describe the so called
domain walls \cite{vachaspati} connecting different portions of the space
were the field is at different values of the degenerate minima of the field
potential. In other words, the field configuration interpolates between two
of those potential minima. At this point, it is important to remark that a
powerful insight to solve nonlinear problems analytically was introduced by
Bogomolnyi, Prasad and Sommerfield \cite{Prasad, bogo}. In this case, the
method shown by Bogomolnyi, Prasad and Sommerfield is now called of BPS
approach, and it is based in obtaining a first-order differential equation
from the energy functional. By using this method, it is possible to find
solutions that minimize the energy of the configuration and that ensure
their stability.

In the context of the field theory it is quite common the appearance of
solitons \cite{Rajaraman}, which are field configurations presenting a
localized and shape-invariant aspect, having a finite energy density as well
as being capable of keeping their shape unaltered after a collision with
another solitons. The presence of those configurations is nowadays well
understood in a wide class of models, presenting or not topological nature.
As examples one can cite the monopoles, textures, strings and kinks \cite%
{Vilenkin}. \

An important feature of a large number of interesting nonlinear models is
the presence of topologically stable configurations, which prevents them
from decaying due to small perturbations. Among other types of nonlinear
field configurations, there is a specially important class of time-dependent
stable solutions, the breathers appearing in the Sine-Gordon like models.
Another time-dependent field configuration whose stability is granted for by
charge conservation are the $Q$-balls \ as baptized by Coleman \cite%
{S.Coleman1} or nontopological solitons \cite{T. D. Lee}. However,
considering the fact that many physical systems interestingly may present a
metastable behavior, a further class of nonlinear systems may present a very
long-living configuration, usually known as oscillons. \ This class of
solutions was discovered in the seventies of the last century by Bogolyubsky
and Makhankov \cite{I.L.Bogolyubsky}, and rediscovered posteriorly by
Gleiser \cite{Gleiser1}. Those solutions, appeared in the study of the
dynamics of first-order phase transitions and bubble nucleation. Since then,
more and more works were dedicated to the study of these objects \cite%
{Gleiser1}-\cite{kast}.

Oscillons are quite general configurations and are found in inflationary
cosmological models \cite{Gleiser1}, in the Abelian-Higgs $U(1)$ models \cite%
{Gleiser8}, in the standard model $SU(2)\times U(1)$ \cite{Graham2}, in
axion models \cite{Kolb}, in expanding universe scenarios \cite{Graham1} and
in systems involving phase transitions \cite{Gleiser2}.

The usual oscillon aspect is typically that of a bell shape which oscillates
sinusoidally in time. Recently, Amin and Shirokoff \cite{Mustafa} have shown
that depending on the intensity of the coupling constant of the
self-interacting scalar field, it is possible to observe oscillons with a
kind of plateau at its top. In fact, they have shown that these new
oscillons are more robust against collapse instabilities in three spatial
dimensions.

At this point it is interesting to remark that Segur and Kruskal \cite%
{Segur1} have shown that the asymptotic expansion do not represent in
general an exact solution for the scalar field, in other words, it simply
represents an asymptotic expansion of first order in $\epsilon $, and it is
not valid at all orders of the expansion. They have also shown that in one
spatial dimension they radiate \cite{Segur1}. In a recent work, the
computation of the emitted radiation of the oscillons was extended to the
case of two and three spatial dimensions \cite{Gyula2}. Another important
result was put forward by Hertzberg \cite{Hertzberg}. In that work he was
able to compute the decaying rate of quantized oscillons, and it was shown
that its quantum rate decay is very distinct of the classical one.

On the other hand, some years ago, Kostelecky and Samuel \cite{kostelecky1}
started to study the problem of the Lorentz and \textit{CPT} \ (charge
conjugation-parity-time reversal) symmetry breaking. This was motivated by
the fact that the superstring theories suggest that Lorentz symmetry should
be violated at higher energies. After that \ seminal work, a theoretical
framework about Lorentz and \textit{CPT }\ symmetry breaking has been
rigorously developed. As an example, the effects on the standard model due
to the \textit{CPT} violation and Lorentz breaking were presented by
Colladay and Kostelecky \cite{Colladay}. Recently, a large amount of works
considering the impact of some kind of Lorentz symmetry breaking \ have
appeared in the literature \cite{Maccione}-\cite{adccc7}. As one another
example, recently Belich \textit{et. al. }\cite{Belich} studied\textit{\ }the%
\textit{\ }Aharonov-Bohm-Casher problem with a nonminimal Lorentz-violating
coupling. In that reference the authors have shown that the
Lorentz-violation is responsible by the lifting of the original degeneracies
in the absence of magnetic field, even for a neutral particle.

By introducing a dimensional reduction procedure to $(1+2)$ dimensions
presented \ in Ref. \cite{Belich3}, Casana, Carvalho and Ferreira applied
the approach to investigate the dimensional reduction of the \textit{CPT}%
-even electromagnetic sector of the standard model extension. Another
important work was presented by Boldo \textit{et al. }\cite{Boldo}, where
the problem of Lorentz symmetry violation gauge theories in connection with
gravity models was analyzed. In a very recent work, Kostelecky and Mewes
\cite{Kostelecky2}, also analyzed the effects of Lorentz violation in
neutrinos.

In recent years, investigations about topological defects in the presence of
Lorentz symmetry violation have been addressed in the literature \cite%
{dutracollaborators}-\cite{dutra-roldao-rafael}. Works have also been done
on monopole and vortices in Lorentz violation scenarios \cite{manoel1}. For
instance, \ in Ref. \cite{manoel1}, a question \ about the Lorentz symmetry
violation on BPS vortices was investigated. In that paper, the Lorentz
violation \ allows a control of the radial extension and of the magnetic
field amplitude of the Abrikosov-Nielsen-Olesen vortices.

In fact, Lorentz invariance is the most fundamental symmetry of the standard
model of particle physics and they have been very well verified in several
experiments. But, it is important to remark that we can not be sure that
this, or any other, symmetry is exact apart from an experimental accuracy.
This affirmation is encouraged due to the fact that there exists some
experimental tests of the Lorentz invariance being carried in low energies,
in other words, energies smaller than $14$ Tev. Thus, from this fact, we can
suspect that at high energies the Lorentz invariance could not be preserved.
As an example, in the string theory there is a possibility that we could be
living in an Universe which is governed by noncommutative coordinates \cite%
{adc1}. In this scenario it was shown in Ref. \cite{adc2} that the Lorentz
invariance is broken.

Furthermore, in a cosmological scenario, the occurrence of high energy
cosmic rays above the Greisen-Zatsepin-Kuzmin (GZK) cutoff \cite{adc3} or
super GZK events, has been found in astrophysical data \cite{adc4}. This
event indicate the possibility of a Lorentz violation \cite{adc5}.

The impact of Lorentz violation on the cosmological scenario is very
important, because several of its weaknesses could be easily explained by
the Lorentz violation. For instance, it was shown by Bekenstein \cite{adc6}
that the problem of the dark matter is associated with the Lorentz violating
gravity and in Ref.\cite{adc7} Lorentz violation also is used to clarify the
dark energy problem. Nowadays, the breaking of the Lorentz symmetry is a
fabulous mechanism for description of several problems and conflicts in
cosmology, such as the baryogenesis, primordial magnetic field,
nucleosynthesis and cosmic rays \cite{adc11}.

In the inflationary scenario with Lorentz violation, Kanno and Soda \cite%
{adc12} have shown that Lorentz violation affects the dynamics of the
inflationary model. In this case, that authors showed that, using a
scalar-vector-tensor theory with Lorentz violation, the exact Lorentz
violation inflationary solutions are found in the absence of the inflaton
potential. Therefore, the inflation can be connected with the Lorentz
violation.

Here, it is convenient to us to emphasize that the inflation is the
fundamental ingredient to solve both the horizon as the flatness problems of
the standard model of the very early universe. Approximately $10^{-33}$
seconds after the inflation, the inflaton decays to radiation, where quarks,
leptons and photons were coupled to each other. In this case, the baryonic
matter was prevented from forming. Therefore, approximately $1.388\times
10^{12}$ seconds after the Big Bang, the universe has cooled enough to allow
photons to freely travel through the universe. After that, matter has became
dominant in the universe.

At this point, it is important to remark that the post-inflationary universe
is governed by real scalar fields where nonlinear interactions are present.
Thus, it was shown in Ref. \cite{adc13} that oscillons can easily dominate
the post-inflationary universe. In that work, it was demonstrated that the
post-inflationary universe can contain an effective matter-dominated phase,
during which it is dominated by localized concentrations of scalar field
matter. Furthermore, in a very recent work \cite{adc14}, a class of
inflationary models was introduced, giving rise to oscillons configurations.
In this case, it was argued that these oscillons, could dominate the matter
density of the universe for a given time. Thus, one could naturally wonder
about the effect of Lorentz violation over this scenario.

Thus, in this work we are interested in answer the following issues: Can
oscillons and breathers exist in scenarios with Lorentz violation symmetry?
If oscillons and breathers exists in these scenarios how their profile is
changed? Furthermore, what happens with the lifetime of the oscillons?

Therefore, in this paper, we will show that oscillons and breathers can be
found in Lorentz violation scenarios, our study is performed by using
Lorentz violation theories rigorously derived in the literature \cite%
{Colladay, kostelecky5}. As a consequence, the principal goal here is to
analyze the case of two nonlinearly coupled scalar fields case. However, we
use a constructive approach, so that we start by studying the cases of one
scalar field models and, then, use those results in the study we are
primarily interested in.

This paper is organized as follows. In section 2 we present the description
of the Lagrangian density for a real scalar field in presence of a Lorentz
violation scenario. In section 3 we calculate the respective commutation
relations of the Poincaré group in the Standard-Model Extension (SME) in a $%
1+1$ dimensional flat Minkowski space-time. The approach of the equation of
motion is given in section 4. Usual oscillons in the background of the
Lorentz violation is analyzed in section 5. In section 6 we will find the
flat-top oscillons which violates the Lorentz symmetry. The breathers
solutions are presented in section 7. We discuss the outgoing radiation by
oscillons in section 8. In the section 9 we will present the oscillons in a
two scalar field theory. Finally, we summarize our conclusions in section 10.

\section{2. Standard-Model Extension Lagrangian}

In this section, we present a scalar field theory in a $3+1$-dimensional
flat Minkowski space-time, but here we consider a break of the Lorentz
symmetry. In low energy, Lorentz and \textit{CPT }symmetries the standard
model (SM) of particle physics is experimentally well supported, but in high
energies the superstring theories suggest that Lorentz symmetry should be
violated, in this context, the framework to study Lorentz and \textit{CPT}
violation is the so-called standard-model extension. In the description of
the SME, the Lagrangian density for a real scalar field containing Lorentz
violation (LV), which can be read as a simplified version of the Higgs
model, is given by \cite{Colladay, kostelecky5}

\begin{equation}
\mathcal{L}=\frac{1}{2}\partial _{\mu }\varphi \partial ^{\mu }\varphi +%
\frac{1}{2}k^{\mu \nu }\partial _{\mu }\varphi \partial _{\nu }\varphi
-V(\varphi ),  \label{1.11}
\end{equation}

\noindent where $\varphi $ is a real scalar field, $k^{\mu \nu }$ is a
dimensionless tensor which controls the degree of Lorentz violation and $%
V(\varphi )$ is the self-interaction potential. It is important to remark
that, some years ago \cite{dutracollaborators}, this Lagrangian density was
used to study defect structures in Lorentz and \textit{CPT} violating
scenarios. In that case the authors showed that the violation of Lorentz and
\textit{CPT} symmetries is responsible by the appearance of an asymmetry
between defects and antidefects. This was generalized in \cite%
{dutracollaborators}. Furthermore, one similar Lagrangian density have been
applied in the study on the renormalization of the scalar and Yukawa field
theories with Lorentz violation. In that case, it was shown that a LV theory
with $N$ scalar fields, interacting through a $\phi ^{4}$ interaction, can
be written as%
\begin{eqnarray}
\mathcal{L}_{K} &=&\frac{1}{2}(\partial _{\mu }\varphi _{i})(\partial ^{\mu
}\varphi _{i})+\frac{1}{2}\sum_{i=1}^{N}K_{\mu \nu }^{i}\partial ^{\mu
}\varphi _{i}\partial ^{\mu }\varphi _{i}-\frac{1}{2}\lambda ^{2}\varphi
_{i}^{2}  \notag \\
&&  \notag \\
&&+\sum_{i=1}^{N}u_{i}^{\beta }\varphi _{i}\partial _{\beta }\varphi
_{i}+\sum_{j=1}^{N}\varphi _{i}^{2}v_{j}^{\beta }\partial _{\beta }\varphi
_{j}-\frac{g}{4!}(\varphi _{i}^{2})^{2}.
\end{eqnarray}

As a simple example, that authors showed for $K_{\mu \nu
}^{i}=K_{00}^{i}\delta _{\mu }^{0}\delta _{\nu }^{0}$ that the dispersion
relation is given by $E=\sqrt{p^{2}-K_{00}^{i}(p^{0})^{2}+\lambda ^{2}}$,
which implies in a LV. Therefore, using explicit calculations, the quantum
corrections in the above LV theory was studied, and these results show that
the theory is renormalizable.

Now, returning to the equation (\ref{1.11}), we can write the Lagrangian
density in the form

\begin{equation}
\mathcal{L}=\frac{1}{2}(\eta ^{\mu \nu }+k^{\mu \nu })\partial _{\mu
}\varphi \partial _{\nu }\varphi -V(\varphi ).
\end{equation}

In this case, the Minkowsky metric is modified from $g^{\mu \nu }$ to $\eta
^{\mu \nu }+k^{\mu \nu }$, which is responsible for the breaking of the
Lorentz symmetry \cite{Colladay, kostelecky5, Potting}. At this point it is
possible to apply an appropriate linear transformation of the space-time
variable $x^{\mu }$, in order to map the above Lagrangian density into a
Lorentz-like covariant form, but this leads to changes in the fields and
coupling constants of the potential. Thus, the coupling constants and the
fields are rescaled in function of the $k^{\mu \nu }$ parameters. \qquad
\qquad\ \ \

Clearly, as a final product, the LV and Lorentz invariant Lagrangians have
the same equation of motion. The fundamental difference between these two
equations comes from the fact that the new variables $x^{\mu }$ carry
information of the Lorentz violations through of the $k^{\mu \nu }$
parameters. In other words, in the transformed variables, the system looks
to be covariant (under boosts of the transformed space-time variables).
However, as a consequence of the fact that the resulting couplings become
not invariant when one changes from a reference frame to another, there is
no real Lorentz invariance. For instance, such behavior would be analogous
to a change of the value of the electrical charge when one moves from an
inertial reference frame to another one, which is forbidden.

In the Lagrangian density (\ref{1.11}), $k^{\mu \nu }$ is a constant tensor
represented by a $4\times 4$ matrix. It is the term which can be responsible
for the breaking of the Lorentz symmetry. Thus, we write the tensor $k^{\mu
\nu }$ in the form
\begin{equation}
k^{\mu \nu }=\left(
\begin{array}{cccc}
k^{00} & k^{01} & k^{02} & k^{03} \\
k^{10} & k^{11} & k^{12} & k^{13} \\
k^{20} & k^{21} & k^{22} & k^{23} \\
k^{30} & k^{31} & k^{32} & k^{33}%
\end{array}%
\right) ,  \label{2}
\end{equation}

In general $k^{\mu \nu }$ has arbitrary parameters, but it is important to
remark that if this matrix is real, symmetric, and traceless, the \textit{%
CPT }symmetry is kept \cite{Colladay, kostelecky5}. Here, we comment that
under \textit{CPT} operation, $\partial _{\mu }\rightarrow -\partial _{\mu }$%
, the term $k^{\mu \nu }\partial _{\mu }\varphi \partial _{\nu }\varphi $
goes as $k^{\mu \nu }\partial _{\mu }\varphi \partial _{\nu }\varphi
\rightarrow +k^{\mu \nu }\partial _{\mu }\varphi \partial _{\nu }\varphi $.
Thus, one notices that $k^{\mu \nu }$ is always \textit{CPT}-even,
regardless its properties. Furthermore, the tensor $k^{\mu \nu }$ should be
symmetric in order to avoid a vanishing contribution.

In a recent work, Anacleto \textit{et al. }\cite{anacleto} also\textit{\ }%
analyzed a similar process to break the Lorentz symmetry, where the tensor $%
k^{\mu \nu }$ was used to study the problem of acoustic black holes in the
Abelian Higgs model with Lorentz symmetry breaking. In another work by
Anacleto \textit{et al}. \cite{anacleto} the tensor $k^{\mu \nu }$ was used
to study the superresonance effect from a rotating acoustic black hole with
Lorentz symmetry breaking. Finally, in a very recent work \cite{PRD
dutracouceiro}, it was introduced a generalized two-fields model in $1+1$
dimensions which presents a constant tensor and vector functions. In that
case, it was found a class of traveling solitons in Lorentz and \textit{CPT }%
breaking systems.

However, we can to find systems with Lorentz symmetry break which has an
additional scalar field \cite{Potting}%
\begin{eqnarray}
\mathcal{L} &=&\frac{1}{2}\partial _{\mu }\varphi _{1}\partial ^{\mu
}\varphi _{1}+\frac{1}{2}\partial _{\mu }\varphi _{2}\partial ^{\mu }\varphi
_{2}+\frac{1}{2}k^{\mu \nu }\partial _{\mu }\varphi _{1}\partial _{\nu
}\varphi _{1}  \label{pt1} \\
&&  \notag \\
&&-\frac{m_{1}\varphi _{1}^{2}}{2}-\frac{m_{2}\varphi _{2}^{2}}{2}-V(\varphi
_{1},\varphi _{2}).  \notag
\end{eqnarray}

In the above Lagrangian density, we have a different coefficient correcting
the metric, but the coefficients for Lorentz violation cannot be removed
from the Lagrangian density using variables or fields redefinitions and
observable effects of the Lorentz symmetry break can be detected in the
above theory. Therefore, theories with fewer fields and fewer interactions
allow more redefinitions and observable effects.

\section{3. SME Lagrangian: One Field Theory (OFT)}

In this section, we will work in a $1+1$-dimensional Minkowski space-time.
Here, we study a scalar field theory in the presence of a Lorentz violating
scenario. The theory that we will study is given by the Lagrangian density (%
\ref{1.11}). Thus, in this case, the corresponding Lagrangian density must
\begin{equation}
\mathcal{L}_{1+1}=\frac{1}{2}\alpha _{1}(\partial _{t}\varphi )^{2}-\frac{1}{%
2}\alpha _{2}(\partial _{x}\varphi )^{2}+\frac{1}{2}\alpha _{3}\partial
_{t}\varphi \partial _{x}\varphi -V(\varphi ),  \label{h1.1}
\end{equation}

\noindent where
\begin{eqnarray}
\alpha _{1} &\equiv &(1+k^{00}),\,\alpha _{2}\equiv (1-k^{11}),\,\alpha
_{3}\equiv (k^{01}+k^{10}),\,  \notag \\
&& \\
\noindent \partial _{t} &\equiv &\partial /\partial t,\,\partial _{x}\equiv
\partial /\partial x.  \notag
\end{eqnarray}

At this point, it is important to remark that the Lagrangian density clearly
has not manifest covariance. Furthermore, it is possible to observe that the
covariance is recovered by choosing $k^{00}=k^{11}=0$ and $k^{01}=-k^{10}$
(or $k^{01}=k^{10}=0$). Another possibilities that does not represent a LV
are $k^{00}=-k^{11}$ and $k^{01}=-k^{10}$ (or $k^{01}=k^{10}=0$).

Now, from the above, we can easily construct the corresponding Hamiltonian
density%
\begin{equation}
\mathcal{H}=\beta _{1}\Pi ^{2}+\beta _{2}(\partial _{x}\varphi )^{2}+\beta
_{3}\Pi (\partial _{x}\varphi )+V(\varphi ),  \label{h1}
\end{equation}

\noindent where $\beta _{1}=1/(2\alpha _{1})$, $\beta _{2}=[2\alpha
_{1}\alpha _{2}+\alpha _{3}(\alpha _{3}-1)]/(4\alpha _{1})$, $\beta
_{3}=-\alpha _{3}/(2\alpha _{1})$ and $\Pi $ is the conjugate momentum,
which is given by%
\begin{equation}
\Pi =\alpha _{1}\partial _{t}\varphi +(\alpha _{3}/2)\partial _{x}\varphi .
\label{h2}
\end{equation}

Let us now see how the Poincarè algebra is modified in this scenario. The
idea of the present analysis is to see how the Poincaré invariance is
broken. In other words, verify how this scenario has the Lorentz symmetry
violated. Therefore, for this we write down the three Poincarè generators,
the Hamiltonian $H$, the total momentum $P$ and the Lorentz boost $M$%
\begin{eqnarray}
H &=&\int dx\mathcal{H},  \label{h4} \\
&&  \notag \\
P &=&\int dx\left[ \frac{\Pi (\partial _{x}\varphi )}{\alpha _{1}}-\frac{%
\alpha _{3}(\partial _{x}\varphi )^{2}}{2\alpha _{1}}\right] , \\
&&  \notag \\
M &=&\int dx\left\{ t\left[ \frac{\Pi (\partial _{x}\varphi )}{\alpha _{1}}-%
\frac{\alpha _{3}(\partial _{x}\varphi )^{2}}{2\alpha _{1}}\right] -x%
\mathcal{H}\right\} .  \label{h6}
\end{eqnarray}

With this, we can calculate the commutation relations of the Poincarè group.
Thus, after straightforward calculations of the usual commutation relations,
it is not difficult to conclude that%
\begin{eqnarray}
\lbrack H,P] &=&-i\left( \frac{\alpha _{3}}{\alpha _{1}^{2}}\right) \int
dx(\partial _{x}\varphi )(\partial _{x}\Pi ),  \label{h7} \\
&&  \notag \\
\lbrack M,H] &=&-i\int dx\left( (4\beta _{1}\beta _{2}+\beta _{3}^{2})\Pi
(\partial _{x}\varphi )\right.  \label{h8} \\
&&  \notag \\
&&\left. -\frac{\alpha _{3}}{\alpha _{1}^{2}}(\partial _{x}\varphi
)(\partial _{x}\Pi )+2\beta _{2}\beta _{3}(\partial _{x}\varphi )^{2}+2\beta
_{1}\beta _{3}(\Pi )^{2}\right) ,  \notag \\
&&  \notag \\
\lbrack M,P] &=&-i\frac{H}{\alpha _{1}}+i\frac{\alpha _{3}}{2\alpha _{1}^{2}}%
\int dx\left( \Pi (\partial _{x}\varphi )\right.  \label{h9} \\
&&  \notag \\
&&\left. +x(\partial _{x}\varphi )(\partial _{x}\Pi )+\frac{\beta _{3}}{%
4\beta _{1}}(\partial _{x}\varphi )^{2}\right) .  \notag
\end{eqnarray}

From the above relations, we can see that the Poincarè algebra is not
closed, since that the usual commutations are not recovered. As a
consequence, in this scenario we have one violation of the Lorentz symmetry.
However, it is possible to recover the complete commutation relations by
taking $k^{00}=k^{11}=0$ and $k^{01}=-k^{10}$ (or $k^{01}=k^{10}=0$). For
instance, making $k^{00}=k^{11}=0$ and $k^{01}=-k^{10}$ we have%
\begin{equation}
\lbrack H,P]=0,\,[M,H]=-iP,\,[M,P]=-iH.  \label{h10}
\end{equation}

At this point we can verify that, for the case $k^{00}=-k^{11}$ and $%
k^{01}=-k^{10}$ (or $k^{01}=k^{10}=0$), the commutation relations (\ref{h7}%
)-(\ref{h9}) lead to%
\begin{equation}
\lbrack H,P]=0,\,[M,H]=-i\alpha _{1}P,\,[M,P]=-iH/\alpha _{1}.
\end{equation}

However, in the above case, we can recover the usual Poincarè algebra using
the re-scale $P=\tilde{P}/\alpha _{1}$. Thus, such commutation relations
indicates that there is no LV in this tensor configuration.

In summary, the Lagrangian density (\ref{h1.1}) has explicit dependence on
the parameters $k^{00}$, $k^{11}$, $k^{01}$ and $k^{10}$, which is
responsible for the violation of the Lorentz symmetry. This happens due to
the fact that \ the Poincaré invariance is not preserved, as one can see
from (\ref{h7})-(\ref{h9}).

\section{4. Equation of motion in Lorentz violation scenarios: OFT}

In this section, we will study the equation of motion in the presence of the
scenario with Lorentz violation of the previous section. Here, our aim is to
study the case in the $1+1$ -dimensional Minkowski space-time. As a
consequence, we will study the theory that is governed by the Lagrangian
density (\ref{h1.1}). Consequently, the corresponding classical equation of
motion can be written as%
\begin{equation}
\alpha _{1}\frac{\partial ^{2}\varphi (x,t)}{\partial t^{2}}-\alpha _{2}%
\frac{\partial ^{2}\varphi (x,t)}{\partial x^{2}}+\alpha _{3}\frac{\partial
^{2}\varphi (x,t)}{\partial x\partial t}+V_{\varphi }=0,  \label{mov1}
\end{equation}

\noindent where $V_{\varphi }\equiv \partial V/\partial \varphi $. Note that
the above equation is carrying information about the symmetry breaking of
the theory.

Here, if one applies the transformation involving the Lorentz boost in the
above equation of motion, one gets%
\begin{equation}
q_{1}\frac{\partial \varphi ^{2}(x^{,},t^{,})}{\partial t^{,2}}-q_{2}\frac{%
\partial \varphi ^{2}(x^{,},t^{,})}{\partial x^{,2}}+q_{3}\frac{\partial
\varphi ^{2}(x^{,},t^{,})}{\partial x^{,}\partial t^{,}}+V_{\varphi }=0,
\end{equation}

\noindent where%
\begin{equation}
x^{,}=\gamma (x-vt),t^{,}=\gamma (t-vx/c^{2}),\gamma =1/\sqrt{1-(v/c)^{2}},
\end{equation}

\noindent and%
\begin{eqnarray}
q_{1} &=&\gamma ^{2}\left( \frac{\alpha _{1}c^{2}-\alpha _{2}v^{2}-\alpha
_{3}cv}{c^{4}}\right) ,  \notag \\
&&  \notag \\
q_{2} &=&\gamma ^{2}\left( \frac{-\alpha _{1}v^{2}+\alpha _{2}c^{2}+\alpha
_{3}cv}{c^{2}}\right) , \\
&&  \notag \\
q_{3} &=&\gamma ^{2}\left( \frac{-2v\alpha _{1}c+2c\alpha _{2}v-\alpha
_{3}(c^{2}+v^{2})}{c^{3}}\right) .  \notag
\end{eqnarray}

Following\ the above demonstration, we can see clearly that this equation is
not invariant under boost transformations. For instance, we can conclude
that the possibilities [$k^{00}=-k^{11}$, $k^{01}=-k^{10}$] or [$%
k^{00}=-k^{11}$, $k^{01}=k^{10}=0$] leads to the equations%
\begin{eqnarray}
\frac{\alpha _{1}}{c^{2}}\frac{\partial \varphi ^{2}(x,t)}{\partial t^{2}}%
-\alpha _{1}\frac{\partial \varphi ^{2}(x,t)}{\partial x^{2}}+V_{\varphi }
&=&0, \\
&&  \notag \\
\frac{\alpha _{1}}{c^{2}}\frac{\partial \varphi ^{2}(x^{,},t^{,})}{\partial
t^{,2}}-\alpha _{1}\frac{\partial \varphi ^{2}(x^{,},t^{,})}{\partial x^{,2}}%
+V_{\varphi } &=&0.
\end{eqnarray}

Note that there is no modification of the equations, in other words, the
possibilities [$k^{00}=-k^{11}$, $k^{01}=-k^{10}$] or [$k^{00}=-k^{11}$, $%
k^{01}=k^{10}=0$] does not represent a genuine factor for LV.

In order to solve analytically the differential equation (\ref{mov1}) and
simultaneously keep the breaking of the Lorentz symmetry\textit{, }we must
decouple the equation. For this, we apply the rotation%
\begin{equation}
\left(
\begin{array}{c}
x \\
t%
\end{array}%
\right) =\left(
\begin{array}{cc}
\cos (\theta ) & -\sin (\theta ) \\
\sin (\theta ) & \cos (\theta )%
\end{array}%
\right) \left(
\begin{array}{c}
X \\
T%
\end{array}%
\right) ,  \label{rot}
\end{equation}

\noindent where $\theta $ is an arbitrary rotation angle. Thus, the equation
(\ref{mov1}) in the new variables is rewritten as%
\begin{equation}
h_{1}\frac{\partial ^{2}\varphi (X,T)}{\partial T^{2}}-h_{2}\frac{\partial
^{2}\varphi (X,T)}{\partial X^{2}}+V_{\varphi }=0,
\end{equation}

\noindent with the definitions%
\begin{eqnarray}
\theta &\equiv &-\frac{1}{2}\arctan \left( \frac{\alpha _{3}}{\alpha
_{1}+\alpha _{2}}\right) ,\,  \notag \\
&&  \notag \\
h_{1} &\equiv &\frac{\alpha _{1}^{2}-\alpha _{2}^{2}+[\alpha
_{3}^{2}+(\alpha _{1}+\alpha _{2})^{2}]\cos (2\theta )}{2(\alpha _{1}+\alpha
_{2})}, \\
&&  \notag \\
h_{2} &\equiv &\frac{\alpha _{2}^{2}-\alpha _{1}^{2}+[\alpha
_{3}^{2}+(\alpha _{1}+\alpha _{2})^{2}]\cos (2\theta )}{2(\alpha _{1}+\alpha
_{2})}.  \notag
\end{eqnarray}

Note that the rotation angle $\theta $ has been chosen in order to eliminate
the dependence in the term $\partial ^{2}\varphi /\partial X\partial T$.
Now, performing the dilations $T=\sqrt{h_{1}}\Upsilon $ and $X=\sqrt{h_{2}}Z$%
, one gets%
\begin{equation}
\frac{\partial ^{2}\varphi (Z,\Upsilon )}{\partial \Upsilon ^{2}}-\frac{%
\partial ^{2}\varphi (Z,\Upsilon )}{\partial Z^{2}}+V_{\varphi }=0.
\label{eqmov}
\end{equation}

From now on we will use the above equation to describe the profile of
oscillons and breathers. It is of great importance to remark that the above
equation has all the information about the violation of the Lorentz
symmetry. In fact, the field $\varphi (Z,\Upsilon )$ carries on the
dependence of the parameters that break the Lorentz symmetry, this
information arises from the fact that new variables $Z$ and $\Upsilon $ have
explicit dependence on the $k^{\mu \nu }$ elements.

\section{5. Usual oscillons with Lorentz violation: OFT}

Now, we study the case of a scalar field theory which supports usual
oscillons in the presence of Lorentz violating scenarios. The profile of the
usual oscillons is one in which the spatial structure is localized in the
space and, in the most cases, is governed by a function of the type $\
sech(x)$. On the other hand, the temporal structure is like $\cos (t)$,
which is periodic. The theory that we will study is given by the Lagrangian
density (\ref{h1.1}). In this case, we showed in the last section that the
corresponding classical equation of motion, after some manipulations, can be
represented by the equation (\ref{eqmov}). Thus, in order to analyze usual
oscillons in this situation, we choose the potential that was used in \cite%
{Mustafa}, which is written as%
\begin{equation}
V(\varphi )=\frac{1}{2}\varphi ^{2}-\frac{1}{4}\varphi ^{4}+\frac{g}{6}%
\varphi ^{6},  \label{6}
\end{equation}

\noindent where $g$ represents a free coupling constant and we will consider
a regime where $g>>1$.

Since our primordial interest is to find periodic and localized solutions,
it is useful, as usual in the study of the oscillons, to introduce the
following scale transformations in $t$ and $x$%
\begin{equation}
\tau =\omega \Upsilon ,\qquad y=\epsilon Z,  \label{7}
\end{equation}

\noindent with $\omega =\sqrt{1-\epsilon ^{2}}$. Thus, the equation of the
motion (\ref{eqmov}) becomes%
\begin{equation}
\omega ^{2}\frac{\partial ^{2}\varphi (y,\tau )}{\partial \tau ^{2}}%
-\epsilon ^{2}\frac{\partial ^{2}\varphi (y,\tau )}{\partial y^{2}}+\varphi
-\varphi ^{3}+g\varphi ^{5}=0.  \label{8}
\end{equation}

Now we are in a position to investigate the usual oscillons. But it is
important to remark that the fundamental point is that here we have the
effects of the Lorentz symmetry breaking. We can see this by inspecting the
above equation of motion, which is carrying information about the terms of
the Lorentz breaking through the variables $y$ and $\tau $. We observe that
it is possible to recover the original equation of motion for usual
oscillons choosing $k^{00}=k^{11}=0$ and $k^{01}=-k^{10}$ (or $%
k^{01}=k^{10}=0$). In this case the Lorentz symmetry is recovered.

Next we expand $\varphi $ as
\begin{equation}
\varphi (y,\tau )=\epsilon \varphi _{1}(y,\tau )+\epsilon ^{3}\varphi
_{3}(y,\tau )+\epsilon ^{5}\varphi _{5}(y,\tau )+....  \label{12}
\end{equation}

Note that the above expansion has only odd powers of $\epsilon $, this
occurs because the equation is odd in $\varphi $. Let us now substitute this
expansion of the scalar field into the equation of motion (\ref{8}). This
leads to%
\begin{eqnarray}
&&\left. \frac{\partial ^{2}\varphi _{1}}{\partial \tau ^{2}}+\varphi
_{1}=0,\right.  \label{13} \\
&&  \notag \\
&&\left. \frac{\partial ^{2}\varphi _{3}}{\partial \tau ^{2}}+\varphi _{3}-%
\frac{\partial ^{2}\varphi _{1}}{\partial \tau ^{2}}-\frac{\partial
^{2}\varphi _{1}}{\partial y^{2}}-\varphi _{1}^{3}=0.\right.  \label{14}
\end{eqnarray}

Therefore, the solution of equation (\ref{13}) is of the form%
\begin{equation}
\varphi _{1}(y,\tau )=\Phi (y)\cos (\tau ),  \label{15}
\end{equation}

Here we call attention to the fact that the solution must be smooth at the
origin and vanishing when $y$ becomes infinitely large.

In order to find the solution of $\Phi (y)$, let us substitute the solution
obtained for $\varphi _{1}(y,\tau )$ into the equation (\ref{14}). Thus, it
is not hard to conclude that
\begin{equation}
\frac{\partial ^{2}\varphi _{3}}{\partial \tau ^{2}}+\varphi _{3}=\left(
\frac{d^{2}\Phi }{dy^{2}}-\Phi +\frac{3}{4}\Phi ^{3}\right) \cos (\tau )+%
\frac{1}{4}\Phi ^{3}\cos (3\tau ).
\end{equation}

Solving the above equation we find a term which is linear in the time-like
variable $\tau $, resulting into a non-periodical solution, and we are
interested in solutions which are periodical in time. Then to avoid this we
shall impose that

\begin{equation}
\frac{d^{2}\Phi }{dy^{2}}-\Phi +\frac{3}{4}\Phi ^{3}=0.  \label{17}
\end{equation}

At this point, one can verify that the above equation can be integrated to
give
\begin{equation}
\left( \frac{d\Phi }{dy}\right) ^{2}+U(\Phi )=E,  \label{19}
\end{equation}

\noindent where $U(\Phi )=-\Phi ^{2}+(3/8)\Phi ^{4}$. Note that in the above
equation, the arbitrary constant $E$ should be set to zero in order to get
solitonic solution. This condition allows the field configuration to go
asymptotically to the vacua of the field potential $U(\Phi )$. Now, we must
solve the equation (\ref{19}) with $E=0$. In this case one gets%
\begin{equation}
\Phi (y)=\sqrt[4]{\frac{8}{3}}[\sech (y)]^{1/2}.  \label{21}
\end{equation}

As one can see, up to the order $\mathcal{O}(\epsilon )$, the corresponding
solution for the field in the original variables is given by%
\begin{eqnarray}
\varphi _{osc}(x,t) &=&\epsilon \sqrt[4]{\frac{8}{3}}\left( \sqrt{\sech\left[
\frac{\epsilon \lbrack x\cos (\theta )+t\sin (\theta )}{\sqrt{h_{2}}}\right]
}\right)  \notag \\
&& \\
&&\times \cos \left[ \frac{\omega \lbrack -x\sin (\theta )+t\cos (\theta )]}{%
\sqrt{h_{1}}}\right] +\mathcal{O}(\epsilon ^{3}).  \notag
\end{eqnarray}

The profile of the above solution is plotted in Fig. 1 for some values of
the $k^{\mu \nu }$ parameters. In the Figure 1 we see the profile of the
usual oscillon \ in the presence of the background of the Lorentz breaking
symmetry. In this case, one can check that the dependence of the solution on
the Lorentz breaking parameters is responsible for a kind of deformation of
the configuration, where the field configuration becomes oscillatory in a
localized region near its maximum value. Furthermore, in the course of the
time, it is possible to observe that the Lorentz breaking symmetry produces
a displacement of the oscillon along the spatial direction. In this case we
will call these configurations as "enveloped oscillons", since in $t=0$ the
new configuration is enveloped by the oscillon with Lorentz symmetry.

Moreover, one can note that if one wants to recover the Lorentz symmetry, it
is necessary to impose that $k^{00}=k^{11}=0$ and $k^{01}=-k^{10}$ (or $%
k^{01}=k^{10}=0$).

\section{6. Flat-top oscillons with Lorentz violation: OFT}

Some years ago, a new class of oscillons, which is characterized by a kind
of plateau at its top, was presented by Amin and Shirokoff \cite{Mustafa}.
In that work, the authors have shown that this configuration has an
important impact on an expanding universe. Thus, in this section, we will
describe the impacts of the Lorentz violation over the flat-top oscillons.
We will study the case in $1+1$-dimensional Minkowski space-time where the
classical equation of motion is given by (\ref{eqmov}). Also, in order to
analyze the flat-top oscillons in this scenario, we choose the potential
that was used in \cite{Mustafa}, which is represented in (\ref{6}).

Now, we begin a direct attack to the problem of finding the flat-top
oscillons. Likewise to the procedure presented in \cite{Mustafa}, we
introduce a re-scaled scalar field by $\varphi (Z,\Upsilon )=\phi (y,\tau )/%
\sqrt{g}$, where $Z=\sqrt{g}\,y$, $\tau =\varpi \Upsilon $ and $\varpi =%
\sqrt{1-\alpha ^{2}/g}$. It is important to remark that the constant $\alpha
^{2}$ is responsible by the change in the frequency, its presence comes from
the nonlinear potential. Thus, it is not difficult to conclude that the
classical equation of motion can be rewritten as%
\begin{equation}
(\partial _{\tau }^{2}\phi +\phi )+g^{-1}[-\alpha ^{2}\partial _{\tau
}^{2}\phi -\partial _{y}^{2}\phi -\phi ^{3}+\phi ^{5}]=0.  \label{10.11}
\end{equation}

So, we are in a position to investigate the so-called flat-top oscillons.
But it is important to remark that the fundamental point is that all the
effects of the Lorentz symmetry breaking are present implicitly in the
classical field. Of course, it is possible to recover the original equation
of motion presented by Mustafa \cite{Mustafa} through a suitable choice of $%
k^{\mu \nu }$.

Let us go further on our search for flat-top oscillons. For this, we expand $%
\phi $ as
\begin{equation}
\phi (y,\tau )=\phi _{1}(y,\tau )+g^{-1}\phi _{3}(y,\tau )+....
\label{12.11}
\end{equation}

If we substitute the above expansion of the scalar field into the equation
of motion (\ref{10.11}), and collect the terms in order $\mathcal{O}(1)$ and
$\mathcal{O}(g^{-1})$, we find
\begin{eqnarray}
&&\left. \frac{\partial ^{2}\phi _{1}}{\partial \tau ^{2}}+\phi
_{1}=0,\right.  \label{13.111} \\
&&  \notag \\
&&\left. \frac{\partial ^{2}\phi _{3}}{\partial \tau ^{2}}+\phi _{3}-\alpha
^{2}\frac{\partial ^{2}\phi _{1}}{\partial \tau ^{2}}-\frac{\partial
^{2}\phi _{1}}{\partial y^{2}}-\phi _{1}^{3}+\phi _{1}^{5}=0.\right.
\label{14.11}
\end{eqnarray}

Therefore, the solution of equation (\ref{13.111}) is of the form%
\begin{equation}
\phi _{1}(y,\tau )=\Psi (y)\cos (\tau ),  \label{15.11}
\end{equation}

In order to find the solution of $\Psi (y)$ let us substitute the solution
obtained for $\phi _{1}(y,\tau )$ into the equation (\ref{14.11}). Thus, it
is not hard to conclude that
\begin{eqnarray}
&&\left. \frac{\partial ^{2}\phi _{3}}{\partial \tau ^{2}}+\phi _{3}=\left(
\frac{d^{2}\Psi }{dy^{2}}-\alpha ^{2}\Psi +\frac{3}{4}\Psi ^{3}-\frac{5}{8}%
\Psi ^{5}\right) \cos (\tau )\right.  \notag \\
&&  \label{16.11} \\
&&\left. +\left( \frac{3}{4}\Psi ^{3}-\frac{5}{16}\Psi ^{5}\right) \cos
(3\tau )-\frac{\Psi ^{5}}{16}\cos (5\tau ).\right.  \notag
\end{eqnarray}

\noindent whose solution can be written as

\begin{eqnarray}
&&\left. \phi _{3}(y,\tau )=\frac{1}{8}[4\,G(y)-2\,H(y)+8\,c_{1}]\cos {%
\left( \tau \right) }\right.  \notag \\
&&  \notag \\
&&\left. -H(y)\,\cos {\left( 3\,\tau \right) }+4\,[G(y)\,\tau
+2\,c_{2}\,]\sin {\left( \tau \right) ,}\right.
\end{eqnarray}

\noindent where we defined that $G(y)\equiv \left( \frac{d^{2}\Psi }{dy^{2}}%
-\alpha ^{2}\Psi +\frac{3}{4}\Psi ^{3}-\frac{5}{8}\Psi ^{5}\right) $ and $%
H(y)\equiv \left( \frac{3}{4}\Psi ^{3}-\frac{5}{16}\Psi ^{5}\right) $.
Furthermore, $c_{1}$ and $c_{2}$ are arbitrary integration constants.

Since that the solution of the function $\phi _{3}$ has a term which is
linear in the variable $\tau $, resulting into a non-periodical solution,
and we are interested in solutions which are periodical in time, we shall
impose that $G(y)$ vanishes. As a consequence we get

\begin{equation}
\frac{d^{2}\Psi }{dy^{2}}=\left( \alpha ^{2}\Psi -\frac{3}{4}\Psi ^{3}+\frac{%
5}{8}\Psi ^{5}\right) ,  \label{17.11}
\end{equation}

At this point, one can verify that the above equation has the same profile
of the equation presented in Ref. \cite{Mustafa}. Therefore, this equation
can be integrated to give
\begin{equation}
\frac{1}{2}\left( \frac{d\Psi }{dy}\right) ^{2}+U(\Psi )=E,  \label{19.11}
\end{equation}

\noindent where $U(\Psi )=-(1/2)\alpha ^{2}\Psi ^{2}+(3/16)\Psi
^{4}-(5/48)\Psi ^{6}$. Note that in the above equation, the arbitrary
constant $E$ should be set to zero in order to get solitonic solution. This
condition allows the field configuration to go asymptotically to the vacua
of the field potential $U(\Psi )$. \ On the other hand, it is usual to
impose that the profile of $\Psi (y)$ be smooth at $y=0$, then it is
necessary to make $d\Psi (0)/dy=0$. As a consequence $E=U(\Psi _{0})=0$,
which implies
\begin{equation}
\alpha ^{2}=\frac{3}{8}\Phi _{0}^{2}-\frac{5}{24}\Phi _{0}^{4},
\label{20.11}
\end{equation}

\noindent with $\Psi _{0}\equiv \Psi (0)$. Thus, solving the above equation
in $\Psi _{0}$, we have a critical value $a\leq $ $\alpha _{c}=\sqrt{27/160}$%
. Above this critical value, $\Psi _{0}$ becomes imaginary.

Now, we must solve the equation (\ref{19.11}) with $E=0$. In this case, we
have%
\begin{equation}
\frac{d\Psi }{\sqrt{\alpha ^{2}\Psi ^{2}-\frac{3}{8}\Psi ^{4}+\frac{5}{24}%
\Psi ^{6}}}=dy.  \label{21.11}
\end{equation}

From this it follows that%
\begin{equation}
\Psi (y)=\frac{(u\sqrt[4]{4vu})}{\sqrt{2\sqrt{v}+\cosh [2y\sqrt{uv(\alpha
_{c}^{2}-\alpha ^{2})}]}},
\end{equation}

\noindent where $v=27/[160(\alpha _{c}^{2}-\alpha ^{2})]$ and $u=(v-1)/v$.

As one can see, up to the order $\mathcal{O}(1)$, the corresponding solution
for the field in the original variables is given by%
\begin{eqnarray}
&&\left. \varphi _{FT}(x,t)=\right. \\
&&  \notag \\
&&\frac{u\sqrt[4]{4vu}}{\sqrt{2g\sqrt{v}+g\cosh \left\{ \frac{2[x\cos
(\theta )+t\sin (\theta )]\sqrt{uv(\alpha _{c}^{2}-\alpha ^{2})}}{\sqrt{%
gh_{2}}}\right\} }}  \notag \\
&&  \notag \\
&&\left. \times \cos \left\{ \frac{\varpi \lbrack -x\sin (\theta )+t\cos
(\theta )]}{\sqrt{h_{1}}}\right\} +\mathcal{O}(g^{-3/2}).\right.  \notag
\end{eqnarray}

The profile of the above solution is plotted in Fig. 2. In the Figure 2 we
see the profile of the flat-top oscillon \ in the presence of the background
of the Lorentz breaking symmetry. In this case, one can check that the
dependence of the solution on the Lorentz breaking parameters is responsible
for a control of the size of the oscillon plateau. Thus, by measuring the
width of the oscillon one could be able to verify the existence and the
degree of the breaking of the symmetry. In Fig. 3 we see the typical profile
of the flat-top oscillon.

There one can note that the effect of the Lorentz breaking over the \ energy
density, it is to becoming it more and more localized around the origin.

\section{7. Breathers with Lorentz violation: OFT}

We will now construct the profile of a breather in a $1+1$ dimensional
Minkowski space-time. Again, we will use the classical equation of motion (%
\ref{eqmov}). The breather solutions arise from the sine-Gordon model%
\begin{equation}
V(\varphi )=\frac{\gamma }{\beta }[1-\cos (\beta \varphi ].
\end{equation}

The sine-Gordon model is invariant under $\varphi \rightarrow \varphi +2n\pi
$, where $n$ is an integer number. In this case, the classical equation of
motion is%
\begin{equation}
\frac{\partial ^{2}\varphi (Z,\Upsilon )}{\partial \Upsilon ^{2}}-\frac{%
\partial ^{2}\varphi (Z,\Upsilon )}{\partial Z^{2}}+\gamma \sin (\beta
\varphi )=0.
\end{equation}

The above equation can be solved by the inverse-scattering method \cite%
{Gardner}. Thus, after straightforward calculations we conclude that the
breather solution is given by%
\begin{equation}
\varphi _{B}(Z,\Upsilon )=\frac{4}{\beta }\arctan \left[ \frac{\sqrt{\gamma
-w^{2}}\sin (w\Upsilon )}{w\cosh (Z\sqrt{\gamma -w^{2}})}\right] ,
\end{equation}

\noindent where $w$ is the frequency of oscillation and describe different
breathers. In Figures 4 and 5 we show the behavior of the above solution.

\section{8. Radiation of oscillons with Lorentz violation symmetry: OFT}

An important characteristic of the oscillons is its radiation emission. In a
seminal work by Segur and Kruskal \cite{Segur1} it was shown that oscillons
in one spatial dimension decay emitting radiation. Recently, the computation
of the emitted radiation in two and three spatial dimensions was did in \cite%
{Gyula2}. On the other hand, in a recent paper by Hertzberg \cite{Hertzberg}%
, it was found that the quantum radiation is very distinct of the classic
one. It is important to remark that the author has shown that the amplitude
of the classical radiation emitted can be found using the amplitude of the
Fourier transform of the spatial structure of the oscillon.

Thus, in this section, we describe the outgoing radiation in scenarios with
Lorentz violation symmetry. Here, we will establish a method in $1+1$
dimensional Minkowski space-time that allows to compute the classical
radiation of oscillons in scenarios with Lorentz symmetry breaking. This is
done by following the method presented in \cite{Hertzberg}. This method
suggests that we can write the solution of the classical equation of motion
in the following form%
\begin{equation}
\varphi _{sol}(x,t)=\varphi _{osc}(x,t)+\eta (x,t),  \label{r1}
\end{equation}

\noindent where $\varphi _{osc}(x,t)$ is the oscillon solution and $\eta
(x,t)$ represents a small correction. Let us substitute this decomposition
of the scalar field into the equation of motion (\ref{mov1}). This leads to

\begin{eqnarray}
&&\left. \alpha _{1}\frac{\partial ^{2}\varphi _{osc}}{\partial t^{2}}%
-\alpha _{2}\frac{\partial ^{2}\varphi _{osc}}{\partial x^{2}}+\alpha _{3}%
\frac{\partial ^{2}\varphi _{osc}}{\partial x\partial t}+\alpha _{1}\frac{%
\partial ^{2}\eta }{\partial t^{2}}\right.  \notag \\
&& \\
&&\left. -\alpha _{2}\frac{\partial ^{2}\eta }{\partial x^{2}}+\alpha _{3}%
\frac{\partial ^{2}\eta }{\partial x\partial t}+U(\varphi _{osc},\eta
)=0,\right.  \notag
\end{eqnarray}

\noindent where $U(\varphi _{osc},\eta )$ is a function which depends on the
form of $V_{\varphi _{sol}}(\varphi _{sol})$. In order to decouple the above
equation we apply the rotation (\ref{rot}) and the dilations $T=\sqrt{h_{1}}%
\Upsilon $ and $X=\sqrt{h_{2}}Z$. Thus, we find%
\begin{eqnarray}
&&\left. \frac{\partial ^{2}\varphi _{osc}(Z,\Upsilon )}{\partial \Upsilon
^{2}}-\frac{\partial ^{2}\varphi _{osc}(Z,\Upsilon )}{\partial Z^{2}}+\frac{%
\partial ^{2}\eta (Z,\Upsilon )}{\partial \Upsilon ^{2}}\right.
\label{eqmov11} \\
&&  \notag \\
&&\left. -\frac{\partial ^{2}\eta (Z,\Upsilon )}{\partial Z^{2}}+U(\varphi
_{osc},\eta )=0.\right.  \notag
\end{eqnarray}

From the above equation it is possible to find the solution for $\eta
(Z,\Upsilon )$ which carries the dependence on the parameters that break the
Lorentz symmetry. We want to investigate the model given by (\ref{6}), then
we have%
\begin{eqnarray}
&&\left. U(\varphi _{osc},\eta )=\varphi _{osc}+\eta -\varphi
_{osc}^{3}-\eta ^{3}+3\varphi _{osc}^{2}\eta +3\varphi _{osc}\eta ^{2}\right.
\notag \\
&& \\
&&+g(\varphi _{osc}^{5}+\eta ^{5}+10\varphi _{osc}^{2}\eta ^{3}+10\varphi
_{osc}^{3}\eta ^{2}  \notag \\
&&  \notag \\
&&+5\varphi _{osc}\eta ^{4}+5\varphi _{osc}^{4}\eta ).  \notag
\end{eqnarray}

As $\eta $ represents a small correction, we assume that the nonlinear terms
$\eta ^{2}$, $\eta ^{3}$, $\eta ^{4}$, $\eta ^{5}$ and the parametric
driving terms $3\eta \varphi _{osc}^{2}$, $5g\eta \varphi _{osc}^{4}$ can be
neglected. At this point, it is important to remark that the parametric
driven terms were not considered because we are working in an asymptotic
regime where $\varphi _{osc}$ is also small. In this case, the equation (\ref%
{eqmov11}) takes the form%
\begin{equation}
\frac{\partial ^{2}\eta (Z,\Upsilon )}{\partial \Upsilon ^{2}}-\frac{%
\partial ^{2}\eta (Z,\Upsilon )}{\partial Z^{2}}+\eta (Z,\Upsilon
)=-J(Z,\Upsilon ),  \label{eqmov12}
\end{equation}

\noindent where%
\begin{eqnarray}
&&\left. J(Z,\Upsilon )=\frac{\partial ^{2}\varphi _{osc}(Z,\Upsilon )}{%
\partial \Upsilon ^{2}}-\frac{\partial ^{2}\varphi _{osc}(Z,\Upsilon )}{%
\partial Z^{2}}\right.  \label{eqmov13} \\
&&  \notag \\
&&\left. +\varphi _{osc}(Z,\Upsilon )-\varphi _{osc}^{3}(Z,\Upsilon
)+g\varphi _{osc}^{5}(Z,\Upsilon ).\right.  \notag
\end{eqnarray}

We can use the Fourier transform for solving the differential equation (\ref%
{eqmov12}) where $J(Z,\Upsilon )$ acts as a source. With this in mind, we
write down the Fourier integral transforms%
\begin{eqnarray}
\eta (R,w) &=&\frac{1}{\sqrt{2\pi }}\int dZ\,d\Upsilon \,\eta (Z,\Upsilon )
\label{rad1.11} \\
&&  \notag \\
&&\times \exp [-i(R\,Z-w\Upsilon )],  \notag \\
&&  \notag \\
J(R,w) &=&\frac{1}{\sqrt{2\pi }}\int dZ\,d\Upsilon \,J(Z,\Upsilon )
\label{rad1.12} \\
&&  \notag \\
&&\times \exp [-i(R\,Z-w\Upsilon )].  \notag
\end{eqnarray}

Then, we have the corresponding solution%
\begin{eqnarray}
\eta (Z,\Upsilon ) &=&\frac{1}{\sqrt{2\pi }}\int dR\,dw\,\eta (R,w)
\label{rad1.1} \\
&&  \notag \\
&&\times \exp [i(R\,Z-w\Upsilon )],  \notag
\end{eqnarray}

\noindent where%
\begin{equation}
\eta (R,w)=-\frac{J(R,w)}{R^{2}-(w^{2}+1)}.  \label{rad1.2}
\end{equation}

From the above approach it is possible to find the radiation field for the
oscillons. As a consequence of the method, the oscillons expansion must be
truncated.

\subsection{8.1. SME Usual Oscillons Radiation: OFT}

In this subsection we will study the outgoing radiation of the usual
oscillons in a Lorentz violation scenario. In this case, the oscillon
expansion truncated in order $N$ is given by

\begin{eqnarray}
\varphi (y,\tau ) &=&\epsilon \varphi _{1}(y,\tau )+\epsilon ^{3}\varphi
_{3}(y,\tau )+\epsilon ^{5}\varphi _{5}(y,\tau )  \label{expt} \\
&&  \notag \\
&&+...+\epsilon ^{N}\varphi _{N}(y,\tau ).  \notag
\end{eqnarray}

As an example, we will consider $N=1$. This is the case where the field
configuration corresponds to the oscillon
\begin{equation}
\varphi _{osc}(y,\tau )=\epsilon \,\varphi _{1}(y,\tau ).  \label{rad1.3}
\end{equation}

Substituting (\ref{rad1.3}) in (\ref{eqmov13}), we obtain
\begin{equation}
J(Z,\Upsilon )=\left( \sqrt[4]{\frac{8}{3}}\right) \epsilon ^{3}[\
sech(\epsilon Z)]^{3/2}\cos (3\omega \Upsilon ).
\end{equation}

Thus, for $N=1$ we can solve easily the integral (\ref{rad1.1}) which allows
to find $\eta (Z,\Upsilon )$. Therefore, we can generalize the result to $N$
substituting the expansion (\ref{expt}) in (\ref{eqmov13}), and using the
differential equation (\ref{8}). After the calculations, the result is%
\begin{equation}
J(Z,\Upsilon )=C_{N}\;\epsilon ^{N+2}[\sech(\epsilon Z)]^{N+1/2}\cos (\bar{n}%
\omega \Upsilon )+...,
\end{equation}

\noindent where $C_{N}$ are constant coefficients. For instance, for $N=1$
we have $C_{1}=\sqrt[4]{8/3}$. Next we calculate $\eta (Z,\Upsilon )$ as
given by (\ref{rad1.1}). After straightforward computations, one can
conclude that%
\begin{eqnarray}
&&\left. \eta (Z,\Upsilon )=\frac{\pi \sqrt{\pi }C_{N}\;\epsilon ^{N+2}}{%
k_{rad}}\cos (\omega _{rad}\Upsilon )\right.  \label{rad1.4} \\
&&  \notag \\
&&\left. \times \sin (k_{rad}Z)\int dZ\sech(\epsilon Z)]^{N+1/2}\cos
(k_{rad}Z).\right.  \notag
\end{eqnarray}

\noindent \noindent where%
\begin{equation}
\omega _{rad}=\bar{n}\omega ,\qquad k_{rad}=\sqrt{\omega _{rad}^{2}-1}\text{.%
}
\end{equation}

On the expression (\ref{rad1.4}), we note that there is an outgoing
radiation which has an amplitude described by the integral%
\begin{eqnarray}
A(k_{rad}) &=&\frac{\pi \sqrt{\pi }C_{N}\;\epsilon ^{N+2}}{k_{rad}} \\
&&  \notag \\
&&\times \int dZ\sech(\epsilon Z)]^{N+1/2}\cos (k_{rad}Z),  \notag
\end{eqnarray}

\noindent we also note that the radiation has frequency $\omega _{rad}$ and
wave number $k_{rad}$. We can make use of the above generalization to
calculate the amplitude of radiation of the usual oscillons in Lorentz
violation scenario. For instance, for $N=1$, we have%
\begin{eqnarray}
&&\left. A(k_{rad})=\frac{4\pi \sqrt{2\pi }C_{1}\;\epsilon ^{3}}{k_{rad}}%
\right.  \label{r1.12} \\
&&  \notag \\
&&\left. \times \left[ b_{1}F(a_{1},a_{2},a_{3},-1)+b_{1}^{\ast
}F(a_{1},a_{2}^{\ast },a_{3}^{\ast },-1)\right] ,\right.  \notag
\end{eqnarray}

\noindent where $F(a_{1},a_{2},a_{3},-1)$ and $F(a_{1},a_{2}^{\ast
},a_{3}^{\ast },-1)$ are hypergeometric functions with

\begin{eqnarray}
b_{1} &=&\frac{1}{3\epsilon -2ik_{rad}},\,a_{1}=\frac{3}{2},\, \\
&&  \notag \\
a_{2} &=&\frac{3}{4}-\frac{ik_{rad}}{2\epsilon },\,a_{3}=\frac{7}{4}-\frac{%
ik_{rad}}{2\epsilon }.  \notag
\end{eqnarray}

In Fig. 6 we see how the amplitude of the outgoing radiation changes with
the parameters of $k^{\mu \nu }$. From that Figure one can see that the
amplitude of the outgoing radiation of the oscillons is controlled by the
terms of the Lorentz breaking of the model, in such way that the radiation
amplitude will decay faster when the Lorentz breaking increases.

\subsection{8.2. SME Flat-top oscillons radiation: OFT}

We will now present the outgoing radiation by the Flat-top oscillons in
Lorentz violation scenario. Here, the associated oscillon expansion
truncated in $N$ is defined as

\begin{eqnarray}
&&\left. \varphi (y,\tau )=\varphi _{1}(y,\tau )+\frac{1}{g}\varphi
_{3}(y,\tau )\right.  \label{expt1.3} \\
&&  \notag \\
&&\left. +\frac{1}{g^{2}}\varphi _{5}(y,\tau )+...+\frac{1}{g^{N-1}}\varphi
_{2N-1}(y,\tau ).\right.  \notag
\end{eqnarray}

Substituting the above expansion in (\ref{eqmov13}), we have that
\begin{eqnarray}
&&\left. J(Z,\Upsilon )=\bar{C}_{N}\;\right. \\
&&  \notag \\
&&\left. \times \bar{C}_{N}\left\{ \frac{(u\sqrt[4]{4vu})}{\sqrt{2g\sqrt{v}%
+g\cosh [2Z\sqrt{uv(\alpha _{c}^{2}-\alpha ^{2})/\sqrt{g}}]}}\right\}
^{N+2}\right.  \notag \\
&&  \notag \\
&&\times \cos (\bar{n}\bar{\omega}\Upsilon )+...,  \notag
\end{eqnarray}

\noindent where $\bar{C}_{N}$ are constant coefficients. Now we calculate $%
\eta (Z,\Upsilon )$ as given by (\ref{rad1.1}). After straightforward
computations, one can conclude that%
\begin{eqnarray}
&&\left. \eta (Z,\Upsilon )=\frac{\pi \sqrt{\pi }\bar{C}_{N}\;}{\bar{k}_{rad}%
}\cos (\bar{\omega}_{rad}\Upsilon )\sin (\bar{k}_{rad}Z)\right.
\label{rad4.11} \\
&&  \notag \\
&&\left. \times \int d\tilde{Z}\left\{ \frac{(u\sqrt[4]{4vu})}{\sqrt{2g\sqrt{%
v}+g\cosh [2\tilde{Z}\sqrt{uvg(\alpha _{c}^{2}-\alpha ^{2})}]}}\right\}
^{N+2}\right.  \notag \\
&&  \notag \\
&&\times \cos (\bar{k}_{rad}\tilde{Z}).  \notag
\end{eqnarray}

\noindent \noindent where%
\begin{equation}
\bar{\omega}_{rad}=\bar{n}\bar{\omega},\qquad \bar{k}_{rad}=\sqrt{\bar{\omega%
}_{rad}^{2}-1}\text{.}
\end{equation}

From the above expression, we see that there is an outgoing radiation which
has its amplitude described by the integral%
\begin{eqnarray}
&&\left. A(k_{rad})=\frac{\pi \sqrt{\pi }\bar{C}_{N}\;}{\bar{k}_{rad}}\int
dZ\cos (\bar{k}_{rad}Z)\right. \\
&&  \notag \\
&&\times \left\{ \frac{(u\sqrt[4]{4vu})}{\sqrt{2\sqrt{v}+\cosh [2Z\sqrt{%
uv(\alpha _{c}^{2}-\alpha ^{2})/\sqrt{g}}]}}\right\} ^{N+2}.  \notag
\end{eqnarray}

We can make use the above generalization to calculate the amplitude of
radiation of the Flat-top oscillons in Lorentz violation scenario. For
instance, for $N=1$, we have%
\begin{equation}
A(k_{rad})=\frac{4\pi \bar{C}_{N}\;}{A_{0}\bar{k}_{rad}}\left( \frac{u\sqrt[4%
]{4vu}}{\sqrt{g}}\right) ^{3}(\xi _{1}\mathcal{F}_{a}+\xi _{1}^{\ast }%
\mathcal{F}_{b}),
\end{equation}

\noindent where $\mathcal{F}_{a}=\mathcal{F}(\Omega _{1};\Omega _{2};\Omega
_{2};\Omega _{3},\Omega _{4},\Omega _{5})$ and $\mathcal{F}_{b}=\mathcal{F}%
(\Omega _{1}^{\ast };\Omega _{2};\Omega _{2};\Omega _{3}^{\ast },\Omega
_{4},\Omega _{5})$ are the Appell hypergeometric functions of two variables,
and%
\begin{eqnarray}
A_{0} &=&2\sqrt{\frac{uv(\alpha _{c}^{2}-\alpha ^{2})}{\sqrt{g}}},\,\xi
_{1}=3+\frac{2ik_{rad}}{A_{0}},  \notag \\
&&  \notag \\
\Omega _{1} &=&\frac{3}{2}-\frac{ik_{rad}}{A_{0}},\Omega _{2}=\frac{3}{2}%
,\Omega _{3}=\frac{5}{2}-\frac{ik_{rad}}{A_{0}}, \\
&&  \notag \\
\Omega _{4} &=&\sqrt{A_{0}^{2}-1}-A_{0},\Omega _{5}=\frac{1}{\sqrt{%
A_{0}^{2}-1}-A_{0}}.  \notag
\end{eqnarray}

In this case we see that the amplitude of the outgoing radiation changes
with the parameters $k^{\mu \nu }$. We can see that the amplitude of the
outgoing radiation of the oscillons is controlled by the terms of the
Lorentz breaking of the model, in such way that the radiation amplitude will
decay faster when the Lorentz breaking increases.

\section{9. Oscillons with LV: Two Field Theory (TFT)}

We have seen in section 2 that the most important scenario with LV is that
described by a theory with two scalar fields, because it is possible to find
observable effects of the LV. Then, in this section, we study a two scalar
field theory in the presence of a LV scenario. The theory that we will study
is similar to that given by Potting \cite{Potting}. Here, we will work with
the corresponding Lagrangian density
\begin{eqnarray}
\mathcal{L} &=&\frac{1}{2}\partial _{\mu }\varphi _{1}\partial ^{\mu
}\varphi _{1}+\frac{1}{2}\partial _{\mu }\varphi _{2}\partial ^{\mu }\varphi
_{2}  \label{dlc1} \\
&&  \notag \\
&&+\frac{1}{2}k^{\mu \nu }\partial _{\mu }\varphi _{1}\partial _{\nu
}\varphi _{2}-V\left( \varphi _{1},\varphi _{2}\right) .  \notag
\end{eqnarray}

\noindent where $V\left( \varphi _{1},\varphi _{2}\right) $ is the
interaction potential. For example, in order to find oscillons solutions, we
can to choose the potential in the form%
\begin{eqnarray}
&&\left. V\left( \varphi _{1},\varphi _{2}\right) =\frac{g}{3}\left( \varphi
_{1}^{6}+\varphi _{2}^{6}\right) -\frac{1}{2}\left( \varphi _{1}^{4}+\varphi
_{2}^{4}\right) \right.  \label{dlc2} \\
&&  \notag \\
&&+\varphi _{1}^{2}+\varphi _{2}^{2}+5~g~\left( \varphi _{1}^{4}~\varphi
_{2}^{2}+\varphi _{1}^{2}~\varphi _{2}^{4}\right) -3\varphi _{1}^{2}~\varphi
_{2}^{2}.  \notag
\end{eqnarray}

In order to decouple the Lagrangian density (\ref{dlc1}), we apply the
rotation

\begin{equation}
\left(
\begin{array}{c}
\varphi _{1} \\
\varphi _{2}%
\end{array}%
\right) =\frac{1}{2}\left(
\begin{array}{cc}
1 & 1 \\
1 & -1%
\end{array}%
\right) \left(
\begin{array}{c}
\sigma _{1} \\
\sigma _{2}%
\end{array}%
\right) .  \label{dlc3}
\end{equation}

After straightforward computations, one can conclude that%
\begin{eqnarray}
&&\left. \mathcal{L}=\frac{1}{2}\partial _{\mu }\sigma _{1}\partial ^{\mu
}\sigma _{1}+\frac{1}{2}k_{1}^{\mu \nu }\partial _{\mu }\sigma _{1}\partial
_{\nu }\sigma _{1}\right.  \label{dlc4} \\
&&  \notag \\
&&\left. +\frac{1}{2}\partial _{\mu }\sigma _{2}\partial ^{\mu }\sigma _{2}+%
\frac{1}{2}k_{2}^{\mu \nu }\partial _{\mu }\sigma _{2}\partial _{\nu }\sigma
_{2}-V\left( \sigma _{1},\sigma _{2}\right) ,\right.  \notag
\end{eqnarray}

\noindent where

\begin{equation}
k_{1}^{\mu \nu }=\frac{1}{4}k^{\mu \nu },k_{2}^{\mu \nu }=-\frac{1}{4}k^{\mu
\nu },
\end{equation}

\noindent and the potential is

\begin{equation}
V\left( \sigma _{1},\sigma _{2}\right) =V(\sigma _{1})+V(\sigma _{2}),
\label{dlc5}
\end{equation}

\noindent with

\begin{equation}
V\left( \sigma _{i}\right) =\frac{g}{6}\sigma _{i}^{6}-\frac{1}{4}\sigma
_{i}^{4}+\frac{1}{2}\sigma _{i}^{2},~i=1,2.  \label{dlc6}
\end{equation}

It is important to note that applying the rotations in the fields, the
Lagrangian density was decoupled into two independent Lagrangians $\mathcal{L%
}=\sum\limits_{i=1}^{2}\mathcal{L}_{i}$, where%
\begin{equation}
\mathcal{L}_{i}=\frac{1}{2}\partial _{\mu }\sigma _{i}\partial ^{\mu }\sigma
_{i}+\frac{1}{2}k_{i}^{\mu \nu }\partial _{\mu }\sigma _{i}\partial _{\nu
}\sigma _{i}-V(\sigma _{i}),  \label{dlc7}
\end{equation}

We can see that all the preceding approaches and results can be used here to
find the fields $\sigma _{1}$ and $\sigma _{2}$. Another important point
that it is convenient to remark at this point, comes from the fact that any
variable $x^{\mu }$ redefinition will carry information of the parameter $%
k^{\mu \nu }$ which it is responsible by LV.

As we are working in $1+1$-dimensions, the Lagrangians (\ref{dlc7}) become%
\begin{eqnarray}
\mathcal{L}_{i}^{(1+1)} &=&\frac{1}{2}a_{i}(\partial _{t}\sigma _{i})^{2}-%
\frac{1}{2}b_{i}(\partial _{x}\sigma _{i})^{2}  \label{dlc9} \\
&&  \notag \\
&&+\frac{1}{2}d_{i}\partial _{t}\sigma _{i}\partial _{x}\sigma _{i}-V(\sigma
_{i}),i=1,2.  \notag
\end{eqnarray}

In this case, we have%
\begin{equation}
a_{i}\equiv (1+k_{i}^{00}),\,b_{i}\equiv (1-k_{i}^{11}),\,d_{i}\equiv
(k_{i}^{01}+k_{i}^{10}).
\end{equation}

Now it is quite clear why the Lagrangian density (\ref{dlc1}) is more
important and general than the one described by (\ref{1.11}). First, because
the commutation relations of the Poincarè group is not closed, indicating a
Lorentz Violation. Second, because it is impossible to perform coordinate
changes to eliminate the LV parameters in (\ref{dlc4}), because if we apply
a coordinate change in order to write the Lagrangian in an covariant form,
only one of the sectors will stay invariant.

Now, by using the approaches described in section 4, we find the equations
\begin{equation}
\frac{\partial ^{2}\sigma _{i}(Z_{i},\Upsilon _{i})}{\partial \Upsilon
_{i}^{2}}-\frac{\partial ^{2}\sigma _{i}(Z_{i},\Upsilon _{i})}{\partial
Z_{i}^{2}}+V_{\sigma _{i}}=0,  \label{camp}
\end{equation}

\noindent where
\begin{eqnarray}
Z_{i} &=&\frac{x\cos (\theta _{i})+t\sin (\theta _{i})}{\sqrt{L_{i}}}, \\
&&  \notag \\
\Upsilon _{i} &=&\frac{-x\sin (\theta _{i})+t\cos (\theta _{i})}{\sqrt{H_{i}}%
}.
\end{eqnarray}

\noindent with the set%
\begin{eqnarray}
\theta _{i} &=&-\frac{1}{2}\arctan \left( \frac{d_{i}}{a_{i}+b_{i}}\right) ,
\\
&&  \notag \\
L_{i} &=&\frac{b_{i}^{2}-a_{i}^{2}+[d_{i}^{2}+(a_{i}+b_{i})^{2}]\cos
(2\theta _{i})}{2(a_{i}+b_{i})}, \\
&&  \notag \\
H_{i} &=&\frac{a_{i}^{2}-b_{i}^{2}+[d_{i}^{2}+(a_{i}+b_{i})^{2}]\cos
(2\theta _{i})}{2(a_{i}+b_{i})}.
\end{eqnarray}

Fortunately, we can find periodical solutions for the fields $\sigma _{1}$
and $\sigma _{2}$ from the equation (\ref{camp}). In this case, we are
looking oscillons-like solutions. These solutions were presented in the
sections 5 and 6. Thus, from those sections we can show that%
\begin{eqnarray}
&&\left. \sigma _{i}^{(USUAL)}(x,t)=\right. \\
&&  \notag \\
&&\epsilon _{i}\sqrt[4]{\frac{8}{3}}\left( \sqrt{\sech\left\{ \frac{\epsilon
_{i}[x\cos (\theta _{i})+t\sin (\theta _{i})]}{\sqrt{L_{i}}}\right\} }\right)
\notag \\
&&  \notag \\
&&\left. \times \cos \left\{ \frac{\omega _{i}[-x\sin (\theta _{i})+t\cos
(\theta _{i})]}{\sqrt{H_{i}}}\right\} +\mathcal{O}(\epsilon _{i}^{3}),\right.
\notag
\end{eqnarray}

and

\begin{eqnarray}
&&\left. \sigma _{i}^{(FLAT-TOP)}(x,t)=\right. \\
&&  \notag \\
&&\frac{u_{i}\sqrt[4]{4v_{i}u_{i}}}{\sqrt{2g\sqrt{v_{i}}+g\cosh \left\{
\frac{2[x\cos (\theta _{i})+t\sin (\theta _{i})]\sqrt{u_{i}v_{i}(\alpha
_{c}^{2}-\alpha _{i}^{2})}}{\sqrt{gL_{i}}}\right\} }}  \notag \\
&&  \notag \\
&&\left. \times \cos \left\{ \frac{\varpi _{i}[-x\sin (\theta _{i})+t\cos
(\theta _{i})]}{\sqrt{H_{i}}}\right\} +\mathcal{O}(g^{-3/2}).\right.  \notag
\end{eqnarray}

In the above solutions $\sigma _{i}^{(USUAL)}$ represents the usual
oscillons and $\sigma _{i}^{(FLAT-TOP)}$ are the Flat-Top ones. Furthermore,
we have%
\begin{eqnarray}
\omega _{i} &=&\sqrt{1-\epsilon _{i}^{2}},\varpi _{i}=\sqrt{1-\alpha
_{i}^{2}/g},  \notag \\
&& \\
v_{i} &=&27/[160(\alpha _{c}^{2}-\alpha _{i}^{2})],u_{i}=(v_{i}-1)/v_{i}.
\notag
\end{eqnarray}

As above asserted, the original scalar fields $\varphi _{1}$ and $\varphi
_{2}$ are obtained from the fields $\sigma _{1}$ and $\sigma _{2}$ in the
following form%
\begin{equation}
\varphi _{1}=\frac{\sigma _{1}+\sigma _{2}}{2},\varphi _{2}=\frac{\sigma
_{1}-\sigma _{2}}{2}.
\end{equation}

It is important to remark that the resulting solutions do not present merely
algebraic relation between $\sigma _{i}$ and the original parameters of the
theory, but essentially lead to physical consequences. As one can see, there
are two kind of frequencies which can be combined for each scalar field $%
\varphi _{i}$. This means that their solutions can be considered as a
superposition of two independent fields and, as a consequence, we can have
an interference phenomena in the structure of the oscillon.

An important question concerns the stability of the solutions, given that
each field $\varphi _{i}$ is a combination of the fields $\sigma _{i}$, the stability
and longevity of the oscillons are guaranteed. From a mathematical point of view,
 one can think that the original fields consist of linear combinations
of $\sigma _{i}$. The same occurs when we calculated the outgoing radiation,
 in that case we have two radiation fields $\eta _{1}$ and $\eta _{2}$, which are
independent solutions with small resulting amplitudes. As a consequence,
their linear combinations, $\bar{\eta}_{1}=\eta _{1}+\eta _{2}$ and $\bar{%
\eta}_{2}=\eta _{1}-\eta _{2}$, will give the radiation field of solutions $%
\varphi _{i}$. Therefore, as $\eta _{i}$ are very small solutions, we still
have the stability and longevity of the solutions guaranteed.

\section{10. Conclusions}

In this work we have investigated the so-called flat-top oscillons in the
case of Lorentz breaking scenarios. We have shown that the Lorentz violation
symmetry is responsible for the appearance of a kind of deformation of the
configuration. On the order hand, from inspection of the results coming from
the flat-top oscillons in $1+1$-dimensions with Lorentz breaking in
comparison with the flat-top given in \cite{Mustafa}, one can see that the
oscillons are carrying information about the terms of the Lorentz breaking
of the model, in this case by taking $k^{00}=k^{11}=0$ and $k^{01}=-k^{10}$
(or $k^{01}=k^{10}=0$) one recovers the solution presented in Ref. \cite%
{Mustafa}. Furthermore, this can lead one to obtain the degree of symmetry
breaking by measuring the width of the oscillon in $1+1$ dimensions. One
important question about the non-linear solution is related to its
stability. Thus, we studied the solutions found here by using the procedure
introduced by Hertzberg \cite{Hertzberg,Mustafa}. We concluded that the
radiation emitted by these oscillons is controlled by the terms of the
Lorentz breaking of the model, in such way that the radiation will decay
more quickly as the terms become larger. Finally, all the results obtained
for the case of one scalar field models are promptly extended for the case
of doublets of nounlinearly coupled scalar fields.

Moreover, it is important to highlight that the bounds in Lorentz
violation theories in the Standard Model are very small, and are compatible
with the stability observed for the oscillons here introduced. On the other
hand, observable effects of these oscillons in the real world, are
possible, for instance, in a cosmological context. In that case, the life time of these
oscillons can be decisive in the generation of coherent structures after
cosmic inflation \cite{ddimensional, generation}, where it was shown that oscillons can contribute up
to $20\%$ of the energy density of the Universe. Thus, in this scenario, one
should find bounds on the Lorentz violation which will open a new window to
detect observable effects of breaking Lorentz symmetry. This possibility
is encouraged by the fact that the break of the Lorentz symmetry induces a
kind of beat phenomenon in the structure of the outgoing radiation, in
contrast with the Lorentz invariant case (see Fig. 6). In this
way, in a real world, one can detect the difference in the frequency of the
outgoing radiation, effect that would indicate the presence of a violation of the Lorentz symmetry.
Therefore, in order to dealing with these questions, we are presently working
in a future work where oscillons in cosmological backgrounds with Lorentz
symmetry breaking are presented.

\textbf{Acknowledgments}

The authors thank Professor D. Dalmazi for useful discussions. R. A. C.
Correa thank P. H. R. S. Moraes for discussions regarding cosmology. R. A.
C. Correa also thank Alan Kostelecky for helpful discussions about Lorentz
breaking symmetry. The authors also thank CNPq and CAPES for partial
financial support. Furthermore, the authors also are grateful to the anonymous 
referees for the comments that lead to improving the results and conclusions of this work.

\newpage

\begin{figure*}[tbp]
\centering
\includegraphics[width=16.0cm]{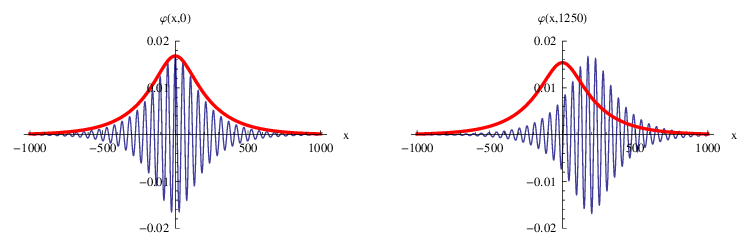}
\caption{Profile of the usual oscillons in $1+1$-dimensions with Lorentz and
\textit{CPT} breaking for $t=0$ (left) and $t=1250$ (right) with $\protect%
\epsilon =0.01$. The thin line corresponds to the case with $k_{00}=0.12$, $%
k_{11}=0.30$, $k_{01}=0.27$ and $k_{10}=0.21$ and the thick line to the case
with $k_{\protect\mu \protect\nu }=0$.}
\label{fig1:oscillons}
\end{figure*}

\begin{figure*}[tbp]
\centering
\includegraphics[width=16.0cm]{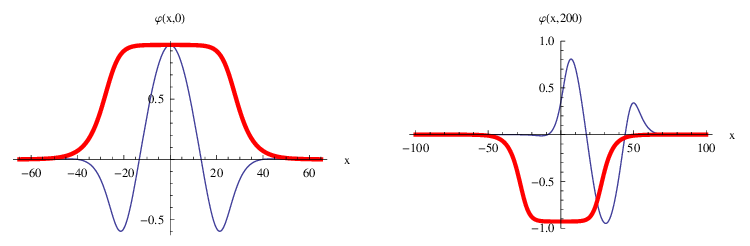}
\caption{Profile of the Flat-Top oscillons in $1+1$-dimensions with Lorentz
symmetry breaking for $t=0$ (left) and $t=200$ (right) with $g=5$. The thin
line corresponds to the case with $k_{00}=0.12$, $k_{11}=0.30$, $k_{01}=0.27$
and $k_{10}=0.21$ and the thick line to the case with $k_{\protect\mu
\protect\nu }=0$. }
\label{fig2:oscillonsscattering}
\end{figure*}

\begin{figure*}[tbp]
\centering
\includegraphics[width=16.0cm]{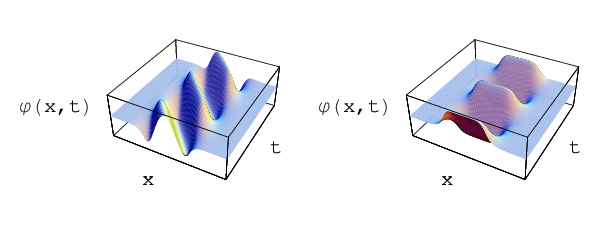}
\caption{Typical profile of the Flat-Top oscillon. The left-hand figure
corresponds to the case with Lorentz breaking symmetry and the right-hand
figure to the one with Lorentz symmetry. }
\label{fig3:oscillons3d}
\end{figure*}

\begin{figure*}[tbp]
\centering
\includegraphics[width=16.0cm]{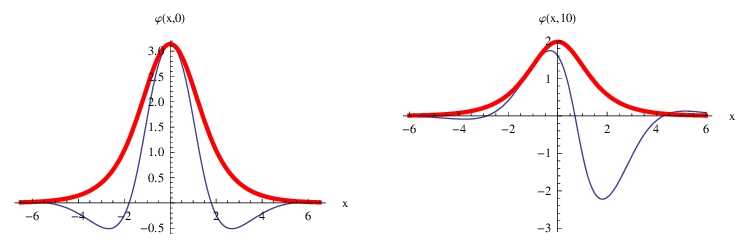}
\caption{Profile of the Breathers $1+1$-dimensions with Lorentz symmetry
breaking for $t=0$ (left) and $t=10$ (right) with $v=2$, $w=1$, $\protect%
\beta =1$. The thin line corresponds to the case with $k_{00}=0.28$, $%
k_{11}=0.30$, $k_{01}=0.27$ and $k_{10}=0.37$ and the thick line to the case
with $k_{\protect\mu \protect\nu }=0$.}
\label{fig4:energydensity}
\end{figure*}

\begin{figure*}[tbp]
\centering
\includegraphics[width=16.0cm]{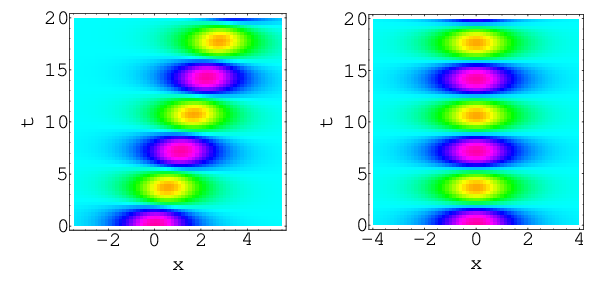}
\caption{Density plot of a Breather. Solution with Lorentz symmetry breaking
(left) and to the one Lorentz symmetry (right).}
\label{fig5:oscillons3+1}
\end{figure*}

\begin{figure*}[tbp]
\centering
\includegraphics[width=16.0cm]{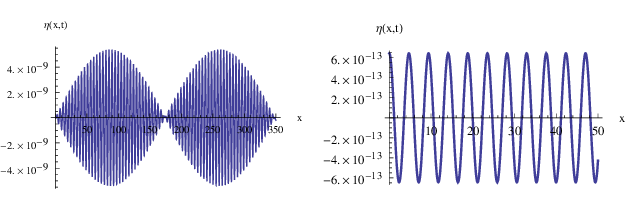}
\caption{Amplitude of the outgoing radiation determined by the Fourier
transform. The left-hand figure corresponds to the case with Lorentz
breaking symmetry and the right-hand to the case with Lorentz symmetry.}
\label{Figure6:Radiation}
\end{figure*}

\end{document}